\PassOptionsToPackage{table,xcdraw}{xcolor}
\documentclass[sigconf]{acmart}

\AtBeginDocument{%
  }

\copyrightyear{2024}
\acmYear{2024}
\setcopyright{rightsretained}
\acmConference[ASEW '24]{39th IEEE/ACM International Conference on
Automated Software Engineering Workshops}{October 27-November 1,
2024}{Sacramento, CA, USA}
\acmBooktitle{39th IEEE/ACM International Conference on Automated Software
Engineering Workshops (ASEW '24), October 27-November 1, 2024, Sacramento,
CA, USA}
\acmDOI{10.1145/3691621.3694949}
\acmISBN {979-8-4007-1249-4/24/10}

\usepackage[table,xcdraw]{xcolor}
\usepackage[framemethod=TikZ]{mdframed}
\usepackage{float}
\usepackage{subcaption}
\usepackage{listings}
\usepackage{stackengine} 
\usepackage{adjustbox}
\definecolor{cyan1}{RGB}{0,255,255}
\definecolor{codegreen}{rgb}{0,0.6,0}
\definecolor{codegray}{rgb}{0.5,0.5,0.5}
\definecolor{codepurple}{rgb}{0.58,0,0.82}
\definecolor{backcolour}{rgb}{0.95,0.95,0.92}

\usepackage{array,multirow}
\usepackage{rotating}
\usepackage{todonotes}
\definecolor{lightblue}{RGB}{200, 220, 255}
\definecolor{vsdarkbackground}{RGB}{30, 30, 30}
\definecolor{vsgreybackground}{RGB}{77,74,60}
\definecolor{vsdarkblue}{RGB}{94, 155, 255}
\definecolor{vsdarkgreen}{RGB}{120, 195, 80}
\definecolor{vsdarkred}{RGB}{255, 85, 85}
\definecolor{vsdarkpurple}{RGB}{177, 97, 211}
\definecolor{vsdarkgray}{RGB}{155, 155, 155}
\definecolor{lightblue}{RGB}{200, 220, 255}
\definecolor{lightgrey}{RGB}{230, 230, 230}
\definecolor{lightorange}{RGB}{255, 201, 102}
\definecolor{lightyellow}{RGB}{255, 255, 0}
\definecolor{loggray}{RGB}{160,160,160}
\definecolor{codegray}{rgb}{0.5,0.5,0.5}
\definecolor{codepurple}{rgb}{0.58,0,0.82}
\definecolor{backcolour}{rgb}{0.95,0.95,0.92}
\definecolor{lightred}{RGB}{255, 110, 110}
\definecolor{lightgreen}{RGB}{181, 230, 29}
\definecolor{codegreen}{rgb}{0,0.6,0}

\newcommand{\stilltodo}[1]{\textcolor{red}{#1}}

\newcommand\oast{\stackMath\mathbin{\stackinset{c}{0ex}{c}{0ex}{\star}{\bigcirc}}}
\makeatother

\lstdefinestyle{mystyle}{   
    commentstyle=\color{codegreen},
    keywordstyle=\color{magenta},
    numberstyle=\tiny\color{codegray},
    stringstyle=\color{codepurple},
    basicstyle=\ttfamily\footnotesize,
    breakatwhitespace=false,         
    breaklines=true,                 
    captionpos=b,   
    frame=single,
    keepspaces=true,                 
    numbers=left,                    
    numbersep=5pt,                  
    showspaces=false,                
    showstringspaces=false,
    showtabs=false,                  
    tabsize=2
}

\begin{document}

\title{Do Android App Developers Accurately Report Collection of Privacy-Related Data?}

\author{Mugdha Khedkar}
\affiliation{%
  \institution{\textit{Heinz Nixdorf Institute \\ Paderborn University}}
  \city{Paderborn}
  \country{Germany}
}
\email{mugdha.khedkar@uni-paderborn.de}

\author{Ambuj Kumar Mondal}
\affiliation{%
  \institution{\textit{Paderborn University}}
  \city{Paderborn}
  \country{Germany}
}
\email{ambuj.mondal@gmail.com}

\author{Eric Bodden}
\affiliation{%
  \institution{\textit{Heinz Nixdorf Institute \\ Paderborn University and Fraunhofer IEM}}
  \city{Paderborn}
  \country{Germany}
}
\email{eric.bodden@uni-paderborn.de}

\begin{abstract}
Many Android applications collect data from users. 
The European Union's General Data Protection Regulation (GDPR) requires vendors to faithfully disclose which data their apps collect. 
This task is complicated because many apps use third-party code for which the same information is not readily available. 
Hence we ask: how accurately do current Android apps fulfill these requirements? 
 
In this work, we first expose a multi-layered definition of privacy-related data to correctly report data collection in Android apps. 
We further create a dataset of privacy-sensitive data classes that may be used as input by an Android app. This dataset takes into account data collected both through the user interface and system APIs. 

We manually examine the data safety sections of 70 Android apps to observe how data collection is reported, identifying instances of over- and under-reporting. 
Additionally, we develop a prototype to statically extract and label privacy-related data collected via app source code, user interfaces, and permissions. 
Comparing the prototype's results with the data safety sections of 20 apps reveals reporting discrepancies. 
Using the results from two \textit{Messaging and Social Media} apps (Signal and Instagram), we discuss how app developers under-report and over-report data collection, respectively, and identify inaccurately reported data categories.  

Our results show that app developers struggle to accurately report data collection, either due to Google's abstract definition of collected data or insufficient existing tool support.
\end{abstract}

\ccsdesc[500]{Security and privacy~Social network security and privacy}
\ccsdesc[300]{Software and its engineering~Software maintenance tools}
\keywords{static analysis, data collection, data protection, privacy-aware reporting}


\maketitle

\section{Introduction}

We use many Android applications in our daily life. The number of apps available in the Google Play Store was most recently placed at 2.43 million, after surpassing 1 million apps in July 2013~\cite{stats}. Among these apps, almost all collect data from their users, and some apps belong to particularly privacy-sensitive categories. 

Since 2018, all Android apps that collect data from users residing in the European Union must comply with the General Data Protection Regulation~\cite{gdpr}. The GDPR aims to protect users' \textit{personal data}, which is defined as \textit{``any information relating to an identified or identifiable natural person"} where an identifiable natural person is one who can be identified, directly or indirectly, by reference to an identifier such as a name, an identification number, etc. The GDPR imposes several obligations on the collection, storage, and processing of personal data. Article 25 of the GDPR highlights \textit{data protection by default and design}, making data protection an integral part of the application development process.

The growing demand for privacy by design~\cite{pbd}, both by end users and by GDPR necessitates that app developers use state-of-the-art technical measures to protect their users’ privacy. 
The legal description of GDPR is very complex and lengthy and hence it can be difficult for app developers to understand. 
Since they lack legal expertise, app developers may find it difficult to understand how to protect users’ data, or even \emph{which} data they should protect~\cite{Horstmann_Domiks_Gutfleisch_Tran_Acar_Moonsamy_Naiakshina_2024}. 

Although many Android applications state a privacy policy, 
violations of privacy policies can easily happen inadvertently, in the worst case just by using data-hungry third-party libraries~\cite{6979855}. 
Moreover, several studies~\cite{privacypolicytrust,automatedriskanalysis,guileak,PTPDroid} have consistently shown that privacy policies diverge significantly from the source code. 
Google Play recently introduced the data safety section, shifting the responsibility of privacy-related reporting to app developers~\cite{data}. They must complete the data safety form, detailing how apps collect, share, and secure users' data.
A recent study by Mozilla~\cite{mozilla} has revealed discrepancies between the information reported in data safety sections and privacy policies of Android apps. 
Thus, it appears that neither documentation provides a usable ground truth in case one wants to study the data collected by Android apps. 
There has been limited research comparing the data safety section with the source code of Android apps, and we aim to fill that gap. 
 
\textbf{Our work.} GDPR's definition of \emph{personal data} brings about a paradigm shift in data protection: apart from protecting data that can cause harm if it is \emph{leaked}, the app must also protect data that can \emph{identify} an individual, either directly or indirectly. 
Our work implements this paradigm shift by centering its data classification around GDPR's definition of \emph{personal data} enabling app developers to conduct a precise risk analysis of the collected user data. 
This is then used to construct a dataset that classifies possible inputs to an Android app using categories similar to those in the data safety section. 
This dataset considers data collected both through the user interface and system APIs, suitable for further static analyses such as taint analysis.

We aim to answer the following research questions:

\textbf{RQ1.} How do data safety sections report data collection and sharing? 

\textbf{RQ2.} How does the reported data collection compare to the data actually collected by the source code?

We manually examine the data safety sections of 70 Android apps to observe how data collection is reported, identifying instances of over- and under-reporting (\textbf{RQ1}). 
Additionally, we develop a prototype to statically extract and label privacy-related data collected via app source code, user interfaces, and permissions.
Comparing the prototype's results with the data safety sections of 20 apps reveals reporting discrepancies. 
Using the results from two \textit{Messaging and Social Media} apps (Signal and Instagram), we discuss how app developers under-report and over-report data collection, respectively, and identify inaccurately reported data categories (\textbf{RQ2}). We make all artifacts available at \href{https://zenodo.org/records/13318828}{https://zenodo.org/records/13318828}.

\textbf{Roadmap.} The remainder of the paper is organized as follows. Section~\ref{background} introduces some background about the GDPR, and data collection in Android apps, and also discusses existing research on the topic. 
Section~\ref{privacyrelevantdata} presents our definition of privacy-related data. 
Section~\ref{approach} discusses how we detect and label the privacy-related data collected by Android apps.  
Section~\ref{experiment} explains our experiment and its findings, and Section~\ref{discussions} discusses our findings in detail. 
Section~\ref{threats} explains the threats to validity, Section~\ref{conclusion} concludes the paper.

\section{Background and Related Work}
\label{background}

This section introduces the concepts we will be discussing in this paper, and discusses existing research on the topic.

\textbf{GDPR and Reporting of Data}. On 25 May 2018, the European Union enforced the General Data Protection Regulation (GDPR) which has been called the \textit{toughest privacy and security law in the world}~\cite{whatisgdpr}. 
The GDPR introduces a strict definition of \textit{personal data} and imposes obligations on the collection, storage, and processing of personal data. 

Article 5 of the GDPR highlights the importance of data minimization, which means collecting only the personal data necessary for specific purposes. 
Once personal data is collected, the GDPR requires app developers to implement technical and organizational measures to protect it. 
These measures include effectively disguising personal data using a \textit{pseudonymization} function. Pseudonymization aims to protect personal data by replacing parts of it with unique, non-identifying \textit{pseudonyms}. 
Personal data can then no longer be attributed to a specific data subject without the use of pseudonyms~\cite{enisa2}. 
Even though pseudonymized data still qualifies as personal data, pseudonymization techniques mitigate data protection risks. 
By pseudonymizing personal data, organizations may continue to use the data for their purposes while reducing the risk of severe consequences in case of a privacy breach~\cite{enisa}. 

Furthermore, Article 13 of the GDPR mandates that personal data collection and processing be reported to the data subject through documents such as privacy policies. 
However, long and convoluted documents may be difficult to understand. 

In 2009, researchers suggested the use of privacy nutrition labels~\cite{2009label1,2009label2} to encourage transparency. 
Following this, the Apple App Store introduced their labels in 2020~\cite{apple}. 
In 2022, researchers introduced a tool~\cite{ioslabelgen} that helps iOS developers generate privacy labels by identifying data flows through code analysis.

\begin{figure*}[ht]
    \centering
    \begin{subfigure}[b]{0.3\textwidth}
         \centering
         \includegraphics[width=0.9\textwidth]{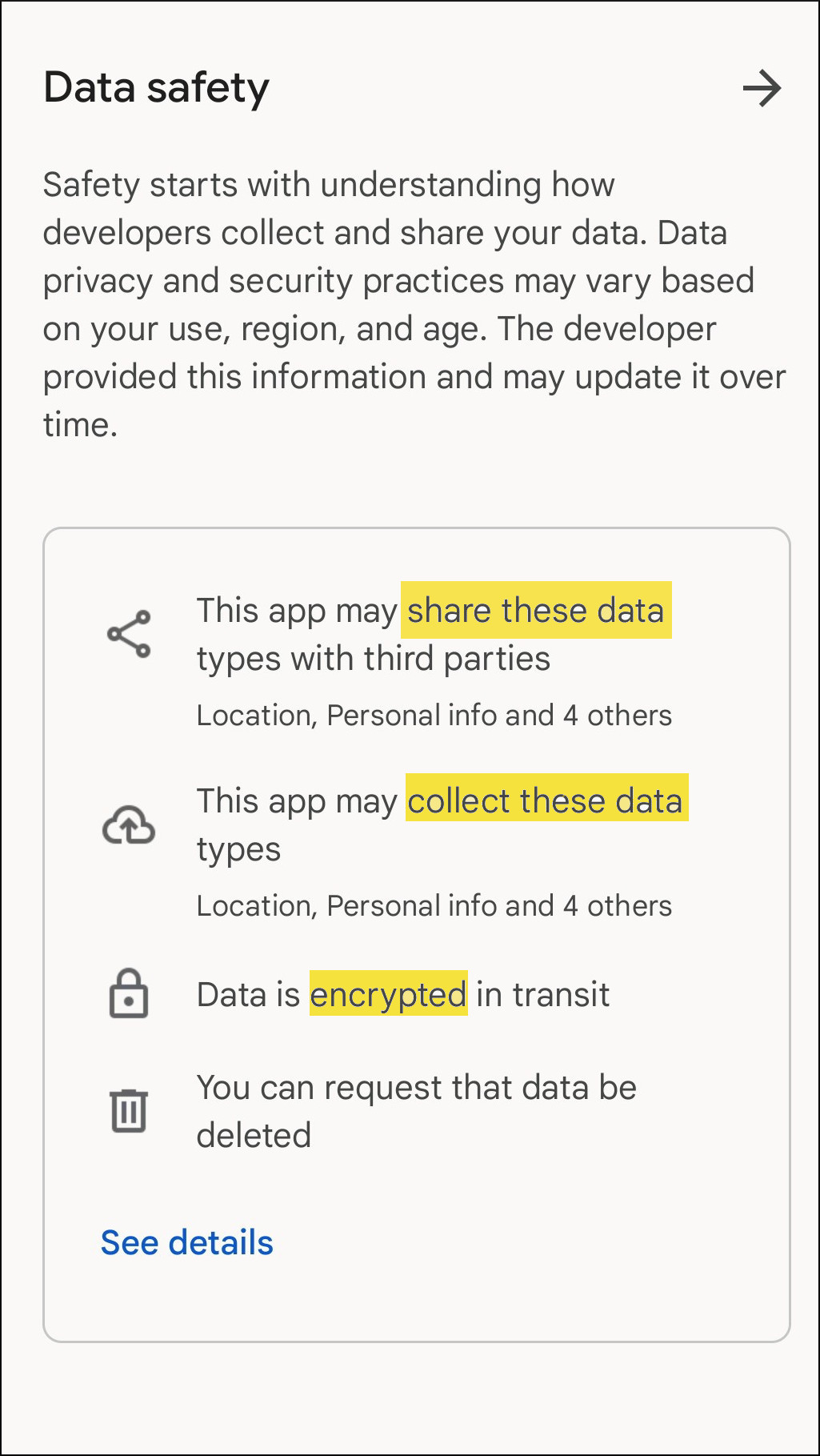}
         \caption{Data safety section}
         \label{fig:dss_1}
     \end{subfigure}
     \hspace{1cm}
     \begin{subfigure}[b]{0.3\textwidth}
         \centering
         \includegraphics[width=0.9\textwidth]{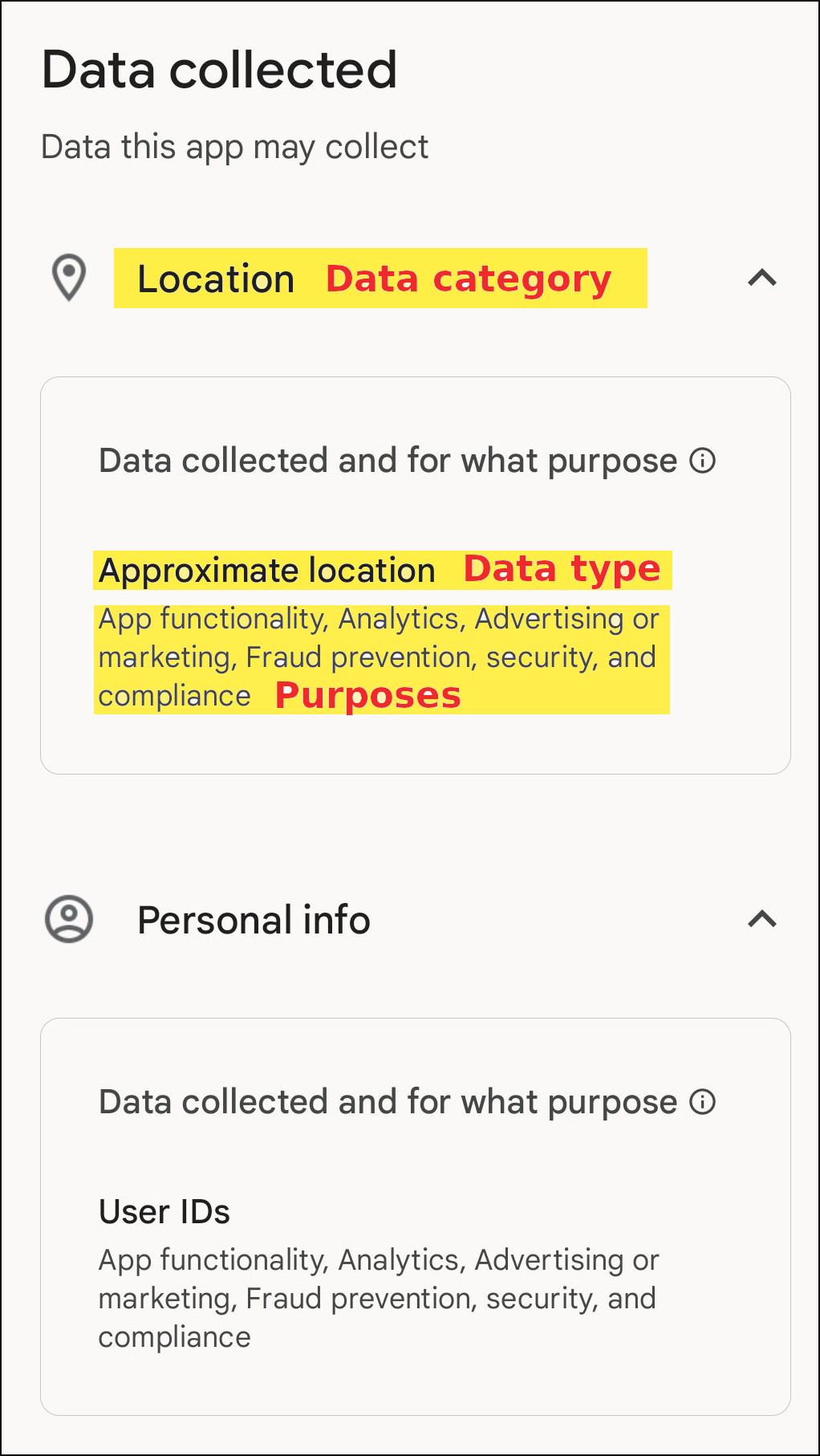}
         \caption{Classification of data}
         \label{fig:dss_2}
     \end{subfigure}
    \centering
    \caption{An example of the data safety section of an Android app.}
    \label{fig:dss}
\end{figure*}

In 2022, Google Play introduced their labels by launching the data safety section~\cite{data}, requiring app developers to disclose how their apps collect, share, and protect user data. 
The data safety form provided by Google comprises three sections: \textit{data sharing, data collection, and security practices} (cf.~Figure~\ref{fig:dss_1}) and is a concise representation of privacy-related information. 
Within this form, user data is classified into different \textit{data categories}, which are further divided into specific \textit{data types} (cf.~Figure~\ref{fig:dss_2}). 
This data can be collected or shared for different \textit{purposes} (chosen from a fixed set). 
App developers are also required to share if the app uses data protection measures such as encryption or on-demand data deletion. 

A recent study by Mozilla~\cite{mozilla} evaluated popular Android apps by grading them based on the presence or absence of discrepancies between their data safety sections and privacy policies. 
Khandelwal et al.~\cite{datalabels} observed the evolution of the data safety section and highlighted under-reporting, over-reporting and inconsistencies in data practices. 
They examined how app developers interact with the data safety sections, using a dataset of snapshots without analyzing user interfaces or source code. 

In 2022, Google introduced Checks~\cite{googlechecks}, a paid service that assists app developers in completing the data safety section. 
However, its cost could hinder many app developers from using its services. 
Recently, open-source alternatives like Matcha~\cite{matcha} and Privado.ai~\cite{privadoai} have been introduced to assist developers in creating accurate Google Play data safety labels. 

\textbf{Data Collection in Android apps.} An Android app may acquire privacy-related data from the user either directly or indirectly. 
The most direct way to gain access to data is to ask the user to provide it via various input screens and user interfaces. 

Android uses XML (eXtensible Markup Language) to describe the user interface layout of an application. 
The XML layout files play a crucial role in defining the buttons, text fields, and layouts that constitute the application's user interface. 
A layout XML file defines the structure and visual appearance of a screen and defines the position, appearance, and characteristics of its components and how they interact with each other. 
\textit{EditText} is a widely used input field that allows users to enter text-based data. It supports various input types, including but not limited to plain text, numbers, email addresses, and passwords. 

\begin{figure}[t]
    \centering
    \includegraphics[width=0.8\linewidth]{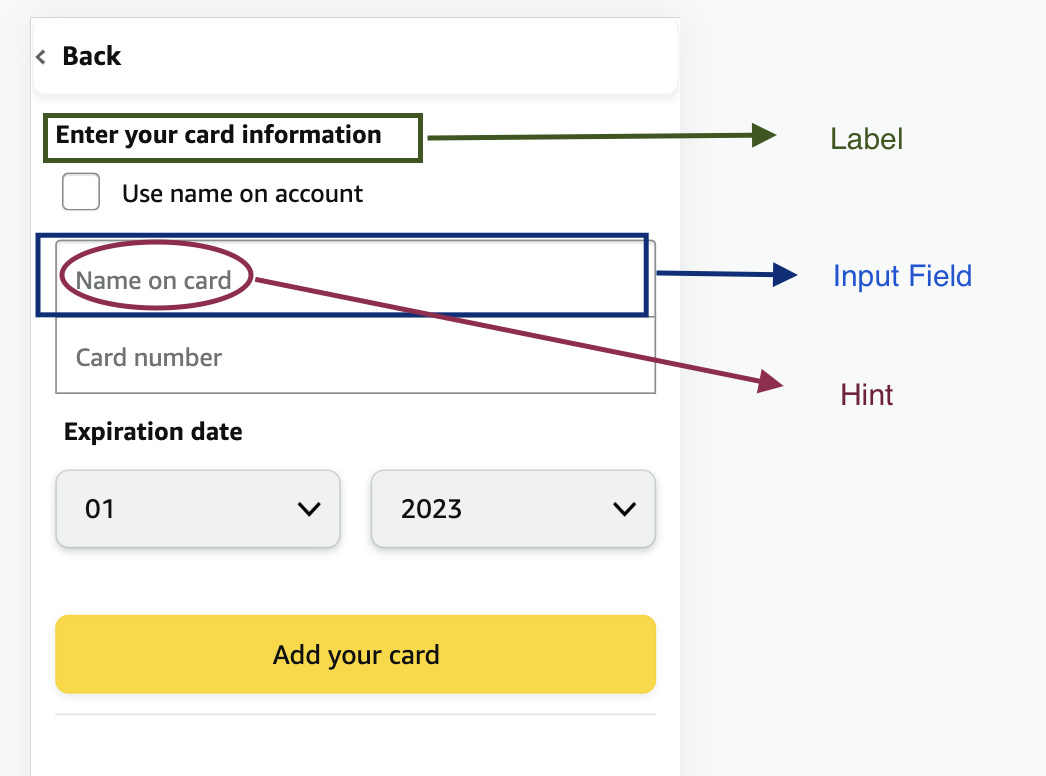}
    \caption{Example: Input fields with labels and Edit Texts}
    \label{fig:labelledInputFields}
\end{figure}

Figure~\ref{fig:labelledInputFields} illustrates a screen from Amazon's shopping app that collects credit card information from the user. 
It consists of \textit{editText} fields for capturing the name and number of the credit card, and dropdown fields to select the expiration date.   
Attributes like \textit{android:inputType} define the allowed input type (e.g., text, number, password), while \textit{android:hint} provides a placeholder or instructional text. 
These attributes describe the type of data to be captured. 
In Figure \ref{fig:labelledInputFields}, using the label and hint texts, one can guess that the input field is asking the user to input credit card details.

Since system calls like \textit{Element.getText()} return the values assigned to these attributes (IDs, hints, labels), static analysis tools~\cite{susi,flowdroid} classify them as sensitive sources. Hence, these tools effectively end up tainting all user inputs that are returned by such sources. 
However, this approach may lack precision, as not all user inputs are equally sensitive, and some may not be sensitive at all. 
Since there are no specific guidelines in API design to regulate this data, it becomes challenging to monitor and identify whether the data actually is privacy-related. 
User interfaces are a common means of directly collecting data from the user, emphasizing the need for a more accurate risk analysis of such user input. 
Several studies analyze the text in user interfaces to detect privacy policy violations and malicious activities~\cite{guileak,AsDroid,whyper}. 
SUPOR~\cite{supor} and UIPicker~\cite{uipicker} can identify sensitive data collected through user interfaces of Android apps. 
Unfortunately, these are in-house proprietary tools and do not focus on GDPR's definition of personal data. 
UiRef~\cite{uiref} analyzes user interfaces to detect sensitive inputs and shows promising results, but its algorithm needs adaptation to the post-GDPR definition of privacy-related data.


An Android app may also acquire data from the operating system through system API calls. 
Some of this data might be privacy-related, such as the GPS information collected through system API calls like \textit{getLastKnownLocation()}. 
Many previous studies on taint analysis~\cite{flowdroid,privflow,iccta} have used predefined sets of methods which are potential sources of system API privacy-related data. 
Follow-up work then proposed machine-learning approaches for automatically classifying and categorizing methods into sources~\cite{susi,swan,swanassist,codoc}. 
In 2023, Kober et al.~\cite{sensitive_data} provided a sound definition of sensitive data derived from the definition of personal data of several legal frameworks (including GDPR). 
They further publicly provided a list of sensitive sources from the Android framework which we used in our work. 

To be able to access certain system APIs in the first place, Android apps may request appropriate permissions from the user at install time or runtime. They are declared in the app's \emph{AndroidManifest.xml} file and relate to various system permissions to read, write or administer the resources. These permissions can be set and reset using the Android settings for the app. 

For the remainder of this paper, we adopt a conservative definition of collected data. This includes data  collected via user interface screens, via system API calls in the source code, or as permissions that Android developer's guide directly maps to specific collected data types in the data safety section~\cite{perm}.

Although related research exists, we did not find approaches that systematically identify and label all collected data as privacy-related categories, allowing direct comparison with the data safety section.

\section{Privacy-related data}
\label{privacyrelevantdata}

The terms ``private'' and ``sensitive'' are subjective, lacking a precise definition of which data is considered sensitive within the context of an Android app~\cite{sensitive_data}. In legal contexts, GDPR's definition of personal data is directly derived from personal identifiers, which are clearly defined by the Article 29 Working Party~\cite{WP29}. These identifiers consist of information that holds a particularly privileged and close relationship with an individual, allowing for identification~\cite{enisa2}. While certain identifiers are sufficient to identify an individual only in some processing contexts, the others may always identify the user. Partial identifiers may only identify users when combined with additional information~\cite{enisa2}. However, until Kober et al.~\cite{sensitive_data} provided a sound definition of sensitive data derived from the definition of personal data of several legal frameworks (including GDPR), there was no such definition for a technical framework such as Android. 
We build on their assessment and expand on their definition of sensitive data, considering the following categories of privacy-related data collected by an Android app: 

\begin{enumerate}
    \item \textit{\colorbox{lightred!40}{Directly identifiable personal data}}: This refers to personal data that can uniquely identify an individual or a device without the use of any additional information. Examples include an email address, a passport number, a phone number, IP address. Such data consists of a single piece of information that is sufficient for identification, and hence it is crucial to pseudonymize it before processing. 
    \item \textit{\colorbox{lightorange!50}{Partially identifiable personal data}}: This type of personal data can identify an individual or a device when combined with additional information. Browsing histories are one example, as they could be linked to some social network profiles such as Twitter or Facebook accounts~\cite{browsing_data}. Kurtz et al.~\cite{fingerprinting} show that users of iOS devices could be singled out through their personalized device configurations, despite the absence of any device identifiers. Hence, this data should be pseudonymized before processing. Other examples include a pin code, postal address, age, and gender. 
    \item \textit{\colorbox{lightyellow!60}{Access data}}: This category includes data that can grant access to an individual's device but cannot identify an individual's device. For example, a password, 
 a card verification value (CVV), and a one-time password (OTP). While it is not categorized as personal data, it still needs protection since it can cause harm if leaked. 
    \item \textit{\colorbox{lightgreen!50}{Context-dependent data}}: This type of data may identify an individual or a device depending on its content and the context in which it is used. Protection is necessary, but its exact identifiability cannot be determined statically. For example, the content of an email or a message may contain privacy-related information that is challenging to identify without running the app.
\end{enumerate}
We summarize these data categories in Table~\ref{tab:data}, assigning a risk ranking to each category. A certain data can get only one risk rank. However, this risk-based classification is too coarse-grained to precisely report data collection in Android apps. To address this issue, we augment data with additional information. To ensure accurate labeling, these augmentations employ a detailed classification and are defined as a pair: Privacy Label $\rightarrow$ Identifier.

\begin{table*}[t]
  \caption{Classification of privacy-related data}
  \label{tab:data}
 \begin{tabular}{|m{5cm}| m{5.75cm}| m{4cm}| m{1cm}|}
    \hline
    \centering
    \textbf{Privacy-Relevance} & \textbf{Categories} & \textbf{Utility}  & 
    \textbf{Risk Rank}\\
    \hline
    \colorbox{lightred!40}{Directly identifiable personal data} & Personal information, Device or other IDs, Financial information & Identification of user &
    1\\\hline
    \colorbox{lightorange!50}{Partially identifiable personal data} & Personal information, Location data, Device data, Audio data, Browsing data, App activity, Photos and videos, Session data, Calendar data & Identification of user possible in combination with other data  & 
    2 \\\hline
    \colorbox{lightyellow!60}{Access data}  & Authentication, Email authentication, Network authentication, Payment authentication & Authentication &
    3 \\\hline
    \colorbox{lightgreen!50}{Context-dependent data} & Message, UI, Audio, Photos and videos, Email  & Unknown statically & 
    4 \\
   \hline
  \end{tabular}
\end{table*}

We label potential data sources (API calls or keywords) with a privacy label, which combines privacy relevance and data category (cf.~Table~\ref{tab:data}). We introduced these data categories by carefully examining the categories in Google's data safety section (cf.~Figure~\ref{fig:dss_2}). We also studied the sensitive Android API sources labeled by SuSi~\cite{susi} and then classified these data categories according to their privacy relevance. In some cases, we introduced some new data categories, such as \colorbox{lightorange!50}{partially identifiable device data}. For every piece of data, we then decided its appropriate category. 

The second component links the data source to an identifier, manually classified based on Android API documentation and Google's data safety section (cf.~Figure~\ref{fig:dss_2}). Table~\ref{tab:examples} gives some concrete examples of privacy-related data labeled according to this classification.

Due to the subjectivity and ambiguity of data identifiers, this classification was done manually. Automating it will be an interesting but non-trivial task, and is beyond the scope of this paper.

\begin{table}[!t]
  \caption{Examples of privacy-related data}
  \label{tab:examples}
 \begin{tabular}{|m{2cm}| m{6cm}|}
    \hline
    \textbf{Data} & \textbf{Privacy label $\rightarrow$ Identifier} \\
    \hline
    Tax ID & {\colorbox{lightred!40}{Directly identifiable financial information} $\rightarrow$ Unique ID} \\
    \hline
     Family name & {\colorbox{lightorange!50}{Partially identifiable personal information} $\rightarrow$ Name} \\
    \hline
    PIN & {\colorbox{lightyellow!60}{Access payment authentication data} $\rightarrow$ Password} \\
    \hline
  \end{tabular}
\end{table}
\section{Approach}
\label{approach}

After establishing a clear definition of privacy-related data, the next task is to detect if Android apps actually collect such data. 
To accomplish this, we first use the data definition presented in Section~\ref{privacyrelevantdata} to construct datasets that classify possible inputs to an Android app (Section~\ref{groundtruth}). 
We then implement a prototype to statically detect and label the collected data (Section~\ref{detectsources}). 

\subsection{Dataset Construction}
\label{groundtruth}

To label sources as potential sources of privacy-related data, we require a dataset that can be used as a classification criterion. 
We constructed two different datasets: an \textit{identifier keywords 
dataset} for labeling UI data elements, and an \textit{identifier API dataset} for labeling system API calls within the app code.

\textbf{Initial corpus.} We referred to Google Play Store and AppBrain statistics~\cite{appbrain} to select the 50 most downloaded apps each from these categories: \textit{Messaging and Social Media, Finance and Banking, News and Entertainment, Sports and Games, Technology and Education, E-Commerce and Shopping}, and \textit{Health and Fitness}. 
In July 2023, we used 
ApkMirror~\cite{apkmirror} to access these 350 apps crawled from the Google Play Store, and manually downloaded these 350 APKs. 

We automatically extracted all the keywords from user input fields along with their metadata (editText id, input type, label, and hint) from the UI layout files of the selected 350 apps. 
To generate the initial dataset of system API methods, we extended SootFX~\cite{sootfx} to detect potential data sources (getter methods) from the source code (APK). 
We first focused on commonly used Android and Java APIs, like \textit{android.location, android.accounts} and \textit{javax.net}. 
We then used the AppBrain statistics~\cite{appbrain} to select the most installed third-party libraries across categories such as analytics and advertisements. 
Table~\ref{tab:tpls} gives some examples of these libraries. 
We extended SootFX to extract potential data sources from the third-party methods found in the source code of the app. 
We combined these with the sources generated by SuSi~\cite{susi} and predefined sources used by FlowDroid~\cite{flowdroid}. 

\textbf{Data annotation.} Once we had the initial corpus for the datasets, we filtered the metadata keywords and manually annotated them to classify them into the data categories described in Section~\ref{privacyrelevantdata}. 
The authors annotated these keywords separately and then discussed their results. 
We conducted a session with 22 research scholars from the Computer Science department of our university, where we explained the data classification and the participants filled out a survey to give their feedback. 
The survey results were considered to update the dataset. 
For example, \textit{last name} was reclassified from \colorbox{lightred!40}{directly identifiable personal information} to \colorbox{lightorange!50}{partially identifiable personal information}. 
We then combined these labeled keywords to generate the \textit{identifier keywords dataset}. A part of the \textit{identifier keywords dataset} is illustrated in Table~\ref{tab:uigroundtruth}. 

The authors manually examined the official API documentation to annotate system API calls from the initial corpus, classifying them into the data categories described in Section~\ref{privacyrelevantdata}. 
In cases where documentation was not available, we studied the labels SuSi~\cite{susi} associated with these calls. 
This annotated information formed the basis for the \textit{identifier API dataset}, as illustrated in Table~\ref{tab:systemgroundtruth}.

We decided to annotate the datasets manually since it was non-trivial to automate this, especially for cases that required extensive discussion among the authors. 
For instance, we categorized IP addresses under \colorbox{lightred!40}{directly identifiable device or other IDs} rather than combining it with other device information and classifying it as \colorbox{lightorange!50}{partially identifiable device data}. 
We carefully discussed ambiguous cases, which could have been classified under multiple categories. 
We classified postal address as \colorbox{lightorange!50}{partially identifiable} \colorbox{lightorange!50}{personal information} and the more general keywords such as zip code, city, or country as \colorbox{lightorange!50}{partially identifiable location data}. 

\begin{table*}[!t]
  \caption{Part of the Identifier API Dataset}
  \label{tab:systemgroundtruth}
 \begin{tabular}{|m{8cm}| m{8.5cm}|}
    \hline
     \textbf{Method signature} & \textbf{Privacy label $\rightarrow$ Identifier} \\
    \hline
     \textit{android.net.IpPrefix: java.net.InetAddress getAddress()} & {\colorbox{lightred!40}{Directly identifiable device or other IDs} $\rightarrow$ IP Address}\\
    \hline
    \textit{android.location.Location: double getLatitude()} & {\colorbox{lightorange!50}{Partially identifiable location data} $\rightarrow$ Approximate location} \\
    \hline
    \textit{com.google.android.gms.auth.api.identity.SignInPassword: java.lang.String getPassword()} & {\colorbox{lightyellow!60}{Access email authentication data} $\rightarrow$ Password} \\
    \hline
    \textit{android.app.Activity: android.view.View findViewById(int)} & \colorbox{lightgreen!50}{Context-dependent UI data} $\rightarrow$ Text field \\
   \hline
  \end{tabular}
\end{table*}

\begin{table}[!t]
  \caption{Part of the Identifier Keywords Dataset}
  \label{tab:uigroundtruth}
 \begin{tabular}{|m{2cm}| m{6cm}|}
    \hline
    \textbf{Keyword(s)} & \textbf{Privacy label $\rightarrow$ Identifier} \\
    \hline
    IBAN, Account number & {\colorbox{lightred!40}{Directly identifiable financial information} $\rightarrow$ Account} \\
    \hline
    First name, Family name & {\colorbox{lightorange!50}{Partially identifiable personal information} $\rightarrow$ Name} \\
    \hline
    PIN, TAN & {\colorbox{lightyellow!60}{Access payment authentication data} $\rightarrow$ Password} \\
    \hline
    Chat & {\colorbox{lightgreen!50}{Context-dependent data} $\rightarrow$ Message} \\
   \hline
  \end{tabular}
\end{table}

\begin{table}[!t]
  \caption{Third-Party Libraries in the Identifier API Dataset}
  \label{tab:tpls}
 \begin{tabular}{m{2.5cm} m{5.5cm}}
 \rowcolor{gray!30}
 \textbf{Category} & \textbf{Library} \\
 Analytics & Google, Firebase \\
 \rowcolor{gray!15} Authentication & Google, Firebase, Apache \\
 Advertisements & Adjust, Applovin, AppsFlyer, Adcolony, Mopub \\
 \rowcolor{gray!15} Image Processing & Google ZXing \\
 Network & okHttp, Apache, gRPC, Volley \\
 \rowcolor{gray!15} Email & Apache James \\
 Location & Baidu \\
 \rowcolor{gray!15} Phone number & Google i18n \\
 \rowcolor{gray!15} validation & \\
 \end{tabular}
 \end{table}

\subsection{Detection of Sources}
\label{detectsources}

\begin{figure}[t]
    \centering 
    \includegraphics[width=0.5\textwidth]{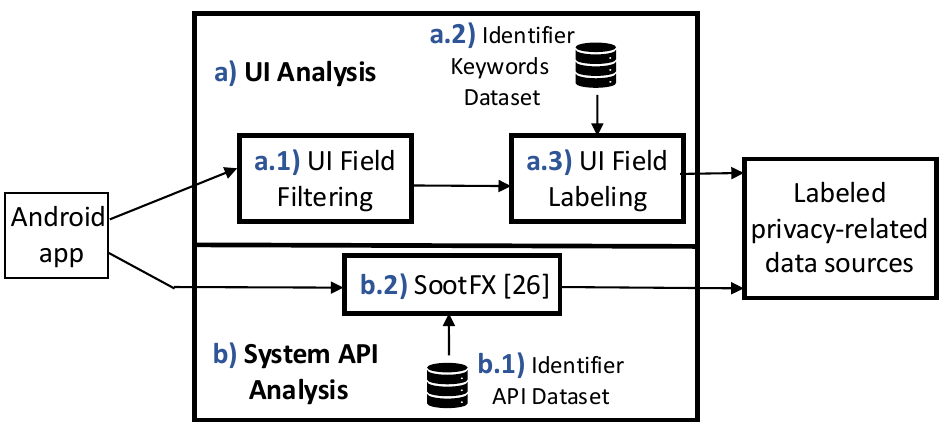}
    \centering
    \caption{Our prototype}
    \label{fig:priceworkflow}
\end{figure}

Our prototype (cf.~Figure~\ref{fig:priceworkflow}) accepts an APK as input and generates labeled privacy-related data sources found in the source code. 
We now discuss the modules needed to detect privacy-related data in the source code.

\textbf{\textcolor[HTML]{4472C4}{a)} UI Analysis.} Analyzing UI screens statically is challenging because the behavior of fields depends on human interaction. 
However, by exploring the attributes associated with these fields, which are defined in the layout XML file for that screen, we can deduce the fields and the data they capture to a certain extent. 

We first decompile the APK file using \textit{apktool}~\cite{apktool} and extract all XML files that present user screens. 
Next, we filter all user input fields and their associated attributes such as identifier, label, hint, and input type (\textcolor[HTML]{4472C4}{\textbf{a.1}}). 
Attributes like \textit{android:inputType} define the allowed input type (e.g., text, number, password), while \textit{android:hint} is an instructional text and suggests the format of data. Attributes like label give additional information about the data to be inserted. 
All \textit{editText} fields have a unique identifier that is used to fetch the value and trigger events. 
It is reasonable to assume that this identifier contains text defining the data it collects (e.g., an identifier like txt\_name suggests a field for collecting a name). 
Combining the identifier with the input type and hint provides a good starting point for heuristics to determine information sensitivity~\cite{supor,uipicker}. 

Our field labeling analysis (\textcolor[HTML]{4472C4}{\textbf{a.3}}) starts by examining the type of the user input field. 
Input fields with types like \textit{Password} are immediately classified as privacy-related information. 
If the input type does not reveal the nature of the information, we check the unique ID of the editText field and compare it with the identifier keywords dataset (\textcolor[HTML]{4472C4}{\textbf{a.2}}). 
In many cases, this unique ID contains important information about the data. 
In case we are still unable to decide the sensitivity, we examine the hint attribute and the text associated with the field. Combining these attributes gives a good idea about the input type. 
We then use the identifier dataset to assign a privacy label to the input field. 
Additionally, we disclose all the permissions declared in the app's \emph{AndroidManifest.xml} file. 
The analysis is currently limited to textual data collected by an Android app. 
Identifying and classifying more complex data (such as images, audio, video, and biometrics) is out of the scope of the analysis. 

\textbf{\textcolor[HTML]{4472C4}{b)} System API Analysis.} To analyze the privacy relevance of system API methods in an Android app, we extend SootFX~\cite{sootfx}, a static code-feature extraction tool for Java and Android. 
SootFX is a standalone tool that uses the Soot framework~\cite{soot} to extract features like methods, classes, and the call graph of the target APK. 
SootFX allows the extraction of whole-program features such as the number of method calls or the number of reachable statements in a target program. 

For our prototype, we introduce a new whole-program feature that uses our \textit{identifier API dataset}  (\textcolor[HTML]{4472C4}{\textbf{b.1}}) as input and checks the APK for privacy-sensitive system API calls. 
To this end, SootFX (\textcolor[HTML]{4472C4}{\textbf{b.2}}) uses the SPARK call graph construction algorithm~\cite{spark} to extract the static call graph of the APK, and matches every method call with the methods described in the \textit{identifier API dataset}. 

If a match is found, it looks for the privacy label corresponding to the method call and labels it as a privacy-related data source.
\section{Experiment}
\label{experiment}

With a functional prototype that statically detects and labels all data collected by an Android app, we aim to answer the following questions:

\textbf{RQ1.} How do data safety sections report data collection and sharing? 

\textbf{RQ2.} How does the reported data collection compare to the data actually collected by the source code?

For our experiment, we initially selected 70 Android apps. In November 2023, we referred to AppBrain statistics~\cite{appbrain} to select 10 
popular apps from each of the following (7) domains: \textit{Messaging and Social Media, 
Banking and Finance, News and Entertainment, Sports and Games, Technology and Education, 
 E-Commerce and Shopping, and Health and Fitness.} 
Based on the findings from RQ1, we narrowed down our focus to 20 apps in RQ2. 
To evaluate RQ2, we used a Linux virtual machine with an Intel(R) Xeon(R) Platinum 8462Y+ 8-core processor with 128 GB memory to run our prototype (Section~\ref{approach}). 
Since the experiment was conducted in November 2023, some data safety sections may have been updated since then. \\

\textit{\textbf{RQ1. How do data safety sections report data collection and sharing?}}

To answer RQ1, we manually examine the data safety sections of 70 apps to understand how data collection and sharing is reported. 
The data safety section classifies all user data into different \textit{data categories}, which are further divided into specific \textit{data types} (cf.~Figure~\ref{fig:dss_2}). 
This data can be collected or shared for 7 different \textit{purposes}.
For RQ1, we classify data safety sections that report the collection of all data types or attribute most data types to 6 or 7 purposes as \textit{over-reporting} practices. 
Conversely, sections that claim no data collection or report only one data type are categorized as \textit{under-reporting} practices.

\textbf{Data collection.} Among all the apps that claim to collect data, only two apps (Signal\footnote{\href{https://signal.org/}{ https://signal.org/}} and Telegram\footnote{\href{https://telegram.org/}{ https://telegram.org/}}) claim not to collect \textit{device or other IDs}. 
The most frequently reported data categories are \textit{device or other IDs, personal information}, and \textit{location}. 
\textit{Sports and Games} apps claim to collect the highest number of data types and categories.  
Surprisingly, 11\% of all data safety forms we analyze \textit{over-report} the data collection, reporting the collection of all data types (Instagram, details below) or attributing most data types to all (7) or nearly all (6) purposes. 
In contrast, 13\% of the forms \textit{under-report} the data collection, either claiming no data collection or reporting the collection of just one data type (Signal, details below). 

\textbf{Data sharing and security practices.} 37\% of the data safety sections we analyze claim that the app does not share any data with third parties, even though their privacy policies indicate otherwise. Western Union\footnote{\href{https://www.westernunion.com}{https://www.westernunion.com/}} claims that data is not collected but is shared. SkyMap\footnote{\href{https://play.google.com/store/apps/details?id=com.google.android.stardroid}{https://play.google.com/store/apps/details?id=com.google.android.stardroid/}} claims that no data is collected or shared, but it can be encrypted. However, the same app claims that users cannot request data to be deleted. We categorize these cases as instances that \textit{inconsistently report} data collection and sharing. \\

\begin{mdframed}[backgroundcolor=black!10,roundcorner=8pt]
\textbf{\textit{Finding 1.}} Many apps appear to inconsistently report data collection and sharing via their data safety sections. 
This is consistent with a large-scale study conducted by Khandelwal et al.~\cite{datalabels}. 
\end{mdframed} 

\textit{\textbf{RQ2. How does the reported data collection compare to the data actually collected by the source code?}}

To answer RQ2, we analyze the app source code and compare its data collection with the data safety section. 
For this experiment, we compare the results of our prototype (Section~\ref{approach}) with the data safety sections of 20 apps across all domains. Due to space constraints, this study focuses on the 20 apps we previously identified as over-reporting and under-reporting data collection (\textbf{RQ1}). 
In addition to detecting privacy-related data in the source code and user interfaces, we statically extract the permissions declared in the app's \emph{AndroidManifest.xml} file. 
Given that the Android developer documentation references app permissions as a means to complete the data safety form~\cite{perm}, we use permissions as another means to examine how effectively these apps report data collection. 

Since the data labels detected by our prototype are similar to the data categories in the data safety form, our experiment allows us to examine 10 distinct data categories out of the 14 present in the data safety section (cf.~Table~\ref{tab:dsscasestudy}). 
Our results validate that \textbf{70\%} of apps reporting \textit{device or other IDs} do indeed collect such data, making it the most consistently reported category, closely followed by \textit{personal information} (\textbf{65\%} apps correctly report it). 
This is encouraging, since these two categories comprise personal data (risks 1 and 2). 
In contrast, \textbf{50\%} of apps reporting \textit{health and fitness data} and \textit{calendar data} do not use the correct manifest permissions, making these the most inconsistently reported data categories. 
We observe that all apps which under-report data collection miss reporting \textit{location} through the data safety section. 
This is concerning since this category encompasses risk 2 personal data. 
We now discuss two apps in detail:

\begin{table}[] 
\caption{Comparison between our results and the data safety section of 20 popular apps. Highlighted apps are discussed in detail.}
\label{tab:dsscasestudy}
\begin{tabular}{lcccccccccc}
\multicolumn{1}{l}{\multirow{2}{*}{\begin{tabular}[c]{@{}c@{}}  \\ \\
\stilltodo{$\star$}: collected, \\ \stilltodo{$\bigcirc$}: reported, \\ $\oast$: collected \\ and \\ reported \\ \end{tabular}}} & \multicolumn{10}{c}{\textbf{Categories in the Data Safety Section}}                                                                          \\ \cline{2-11} 
\multicolumn{1}{l}{}                              & \multicolumn{1}{l}{\textbf{\rotatebox[origin=l]{90}{Device or other IDs}}} & \multicolumn{1}{l}{\textbf{\rotatebox[origin=l]{90}{Personal info}}} & \multicolumn{1}{l}{\textbf{\rotatebox[origin=l]{90}{Audio}}} &  \multicolumn{1}{l}{\textbf{\rotatebox[origin=l]{90}{Contacts}}} &  \multicolumn{1}{l}{\textbf{\rotatebox[origin=l]{90}{Location}}} &  \multicolumn{1}{l}{\textbf{\rotatebox[origin=l]{90}{{Photos and videos}}}} & \multicolumn{1}{l}{\textbf{\rotatebox[origin=l]{90}{{Financial info}}}} & \multicolumn{1}{l}{\textbf{\rotatebox[origin=l]{90}{{Messages}}}} &  \multicolumn{1}{l}{\textbf{\rotatebox[origin=l]{90}{{Health and fitness}}}} & \multicolumn{1}{l}{\textbf{\rotatebox[origin=l]{90}{{Calendar}}}}\\         \midrule
\rowcolor{gray!30}
Signal                                              & \multicolumn{1}{l}{\stilltodo{$\star$}}                                                                        & \multicolumn{1}{l}{$\oast$}                                                                        & \multicolumn{1}{l}{\stilltodo{$\star$}}               & \multicolumn{1}{l}{\stilltodo{$\star$}}                 & \multicolumn{1}{l}{\stilltodo{$\star$}}                  & \multicolumn{1}{l}{\stilltodo{{$\star$}}}                                                                      & \multicolumn{1}{l}{}                                                                          & \multicolumn{1}{l}{\stilltodo{$\star$}}                  & \multicolumn{1}{l}{}                                                                       &    \multicolumn{1}{l}{\stilltodo{$\star$}}                 \\ 
SkyMap                                              & \multicolumn{1}{l}{\stilltodo{$\star$}}                                                                         & \multicolumn{1}{l}{\stilltodo{$\star$}}                                                                        & \multicolumn{1}{l}{}               & \multicolumn{1}{l}{}                  & \multicolumn{1}{l}{\stilltodo{$\star$}}                  & \multicolumn{1}{l}{}                                                                      & \multicolumn{1}{l}{}                                                                          & \multicolumn{1}{l}{}                  & \multicolumn{1}{l}{}                                                                       &         \multicolumn{1}{l}{}          \\ 
Western Union                                       & \multicolumn{1}{l}{\stilltodo{$\star$}}                                                                         & \multicolumn{1}{l}{\stilltodo{$\star$}}                                                                          & \multicolumn{1}{l}{\stilltodo{$\star$}}               & \multicolumn{1}{l}{\stilltodo{$\star$}}                  & \multicolumn{1}{l}{\stilltodo{$\star$}}                  & \multicolumn{1}{l}{\stilltodo{$\star$}}                                                                       & \multicolumn{1}{l}{}                                                                          & \multicolumn{1}{l}{}                  & \multicolumn{1}{l}{}                                                                       &                   \\ 
Yazio                                               & \multicolumn{1}{l}{$\oast$}                                                                        & \multicolumn{1}{l}{$\oast$}                                                                         & \multicolumn{1}{l}{}               & \multicolumn{1}{l}{}                  & \multicolumn{1}{l}{\stilltodo{$\star$}}                  & \multicolumn{1}{l}{$\oast$}                                                                      & \multicolumn{1}{l}{}                                                                          & \multicolumn{1}{l}{}                  & \multicolumn{1}{l}{$\stilltodo{\star}$}                                                                       &                   \\ 
Temu                                                & \multicolumn{1}{l}{$\oast$}                                                                        & \multicolumn{1}{l}{$\oast$}                                                                         & \multicolumn{1}{l}{}               & \multicolumn{1}{l}{}                  & \multicolumn{1}{l}{\stilltodo{$\star$}}                   & \multicolumn{1}{l}{$\oast$}                                                                      & \multicolumn{1}{l}{$\oast$}                                                                          & \multicolumn{1}{l}{}                  & \multicolumn{1}{l}{}                                                                       &                   \\ 
DuckDuckGo                                        & \multicolumn{1}{l}{\stilltodo{$\star$}}                                                                         & \multicolumn{1}{l}{\stilltodo{$\star$}}                                                                          & \multicolumn{1}{l}{\stilltodo{$\star$}}                & \multicolumn{1}{l}{}                  & \multicolumn{1}{l}{\stilltodo{$\star$}}                  & \multicolumn{1}{l}{\stilltodo{$\star$}}                                                                       & \multicolumn{1}{l}{}                                                                          & \multicolumn{1}{l}{}                  & \multicolumn{1}{l}{}                                                                       &                   \\ 
Covpass                                             & \multicolumn{1}{l}{\stilltodo{$\star$}}                                                                         & \multicolumn{1}{l}{\stilltodo{$\star$}}                                                                         & \multicolumn{1}{l}{}               & \multicolumn{1}{l}{}                  & \multicolumn{1}{l}{\stilltodo{$\star$}}                  & \multicolumn{1}{l}{}                                                                      & \multicolumn{1}{l}{}                                                                          & \multicolumn{1}{l}{}                  & \multicolumn{1}{l}{}                                                                       &                   \\ 
Trust                                               & \multicolumn{1}{l}{$\oast$}                                                                        & \multicolumn{1}{l}{$\oast$}                                                                         & \multicolumn{1}{l}{}               & \multicolumn{1}{l}{}                  & \multicolumn{1}{l}{\stilltodo{$\star$}}                   & \multicolumn{1}{l}{}                                                                      & \multicolumn{1}{l}{\stilltodo{$\star$}}                                                                           & \multicolumn{1}{l}{\stilltodo{$\bigcirc$}}                  & \multicolumn{1}{l}{}                                                                       &                   \\ 
Paytm                                               & \multicolumn{1}{l}{\stilltodo{$\star$}}                                                                         & \multicolumn{1}{l}{$\oast$}                                                                         & \multicolumn{1}{l}{\stilltodo{$\star$}}                & \multicolumn{1}{l}{$\oast$}                  & \multicolumn{1}{l}{\stilltodo{$\star$}}                   & \multicolumn{1}{l}{\stilltodo{$\star$}}                                                                      & \multicolumn{1}{l}{\stilltodo{$\star$}}                                                                           & \multicolumn{1}{l}{}                  & \multicolumn{1}{l}{}                                                                       &       \multicolumn{1}{l}{\stilltodo{$\star$}}             \\ 
\rowcolor{gray!30}
Instagram                                           & \multicolumn{1}{l}{$\oast$}                                                                        & \multicolumn{1}{l}{$\oast$}                                                                         & \multicolumn{1}{l}{$\oast$}               & \multicolumn{1}{l}{$\oast$}                  & \multicolumn{1}{l}{$\oast$}                  & \multicolumn{1}{l}{$\oast$}                                                                      & \multicolumn{1}{l}{$\oast$}                                                                          & \multicolumn{1}{l}{\stilltodo{$\bigcirc$}}                  & \multicolumn{1}{l}{\stilltodo{$\bigcirc$}}                                                                     &      \multicolumn{1}{l}{\stilltodo{$\bigcirc$}}              \\ 
Gmail                                               & \multicolumn{1}{l}{$\oast$}                                                                        & \multicolumn{1}{l}{$\oast$}                                                                         & \multicolumn{1}{l}{$\oast$}               & \multicolumn{1}{l}{$\oast$}                  & \multicolumn{1}{l}{$\oast$}                  & \multicolumn{1}{l}{$\oast$}                                                                      & \multicolumn{1}{l}{\stilltodo{$\bigcirc$}}                                                                            & \multicolumn{1}{l}{$\oast$}                  &                \multicolumn{1}{l}{}       &  \multicolumn{1}{l}{$\oast$}                  \\ 
Discord                                             & \multicolumn{1}{l}{$\oast$}                                                                        & \multicolumn{1}{l}{\stilltodo{$\bigcirc$}}                                                                           & \multicolumn{1}{l}{$\oast$}               & \multicolumn{1}{l}{$\oast$}                  & \multicolumn{1}{l}{$\oast$}                  & \multicolumn{1}{l}{$\oast$}                                                                      & \multicolumn{1}{l}{}                                                                          & \multicolumn{1}{l}{\stilltodo{$\bigcirc$}}                   & \multicolumn{1}{l}{}                                                                       &                   \\ 
Youtube                                             & \multicolumn{1}{l}{$\oast$}                                                                        & \multicolumn{1}{l}{$\oast$}                                                                         & \multicolumn{1}{l}{$\oast$}               & \multicolumn{1}{l}{\stilltodo{$\bigcirc$}}                   & \multicolumn{1}{l}{$\oast$}                  & \multicolumn{1}{l}{$\oast$}                                                                      & \multicolumn{1}{l}{$\oast$}                                                                          & \multicolumn{1}{l}{$\oast$}                  & \multicolumn{1}{l}{\stilltodo{$\star$}}                                                                       &           \multicolumn{1}{l}{}        \\ 
Flo                                                 & \multicolumn{1}{l}{$\oast$}                                                                        & \multicolumn{1}{l}{$\oast$}                                                                         & \multicolumn{1}{l}{}               & \multicolumn{1}{l}{}                  & \multicolumn{1}{l}{$\oast$}                  & \multicolumn{1}{l}{\stilltodo{$\bigcirc$}}                                                                       & \multicolumn{1}{l}{$\oast$}                                                                          & \multicolumn{1}{l}{\stilltodo{$\bigcirc$}}                    & \multicolumn{1}{l}{\stilltodo{$\bigcirc$}}                                                                        &             \multicolumn{1}{l}{}      \\ 
\begin{tabular}[c]{@{}l@{}}Candy Crush\\ Saga\end{tabular}                                   & \multicolumn{1}{l}{$\oast$}                                                                        & \multicolumn{1}{l}{$\oast$}                                                                         & \multicolumn{1}{l}{}               & \multicolumn{1}{l}{}                  & \multicolumn{1}{l}{$\oast$}                  & \multicolumn{1}{l}{\stilltodo{$\star$}}                                                                      & \multicolumn{1}{l}{\stilltodo{$\bigcirc$}}                                                                            & \multicolumn{1}{l}{\stilltodo{$\bigcirc$}}                    & \multicolumn{1}{l}{}                                                                       &    \multicolumn{1}{l}{}                \\ 
Babbel                                              & \multicolumn{1}{l}{$\oast$}                                                                        & \multicolumn{1}{l}{\stilltodo{$\bigcirc$}}                                                                           & \multicolumn{1}{l}{$\oast$}               & \multicolumn{1}{l}{}                  & \multicolumn{1}{l}{$\oast$}                  & \multicolumn{1}{l}{\stilltodo{$\star$}}                                                                      & \multicolumn{1}{l}{\stilltodo{$\bigcirc$}}                                                                            & \multicolumn{1}{l}{}                  & \multicolumn{1}{l}{}                                                                       &       \multicolumn{1}{l}{}            \\ 
Google Maps                                         & \multicolumn{1}{l}{$\oast$}                                                                        & \multicolumn{1}{l}{$\oast$}                                                                         & \multicolumn{1}{l}{$\oast$}               & \multicolumn{1}{l}{$\oast$}                  & \multicolumn{1}{l}{$\oast$}                  & \multicolumn{1}{l}{$\oast$}                                                                      & \multicolumn{1}{l}{}                                                                          & \multicolumn{1}{l}{$\oast$}                  & \multicolumn{1}{l}{\stilltodo{$\star$}}                                                                       &                   \\ 
Netflix                                             & \multicolumn{1}{l}{$\oast$}                                                                        & \multicolumn{1}{l}{\stilltodo{$\bigcirc$}}                                                                         & \multicolumn{1}{l}{$\oast$}               & \multicolumn{1}{l}{}                  & \multicolumn{1}{l}{$\oast$}                  & \multicolumn{1}{l}{\stilltodo{$\star$}}                                                                      & \multicolumn{1}{l}{}                                                                          & \multicolumn{1}{l}{}                  & \multicolumn{1}{l}{}                                                                       &     \multicolumn{1}{l}{}              \\ 
Fitbit                                              & \multicolumn{1}{l}{$\oast$}                                                                        & \multicolumn{1}{l}{$\oast$}                                                                         & \multicolumn{1}{l}{\stilltodo{$\star$}}               & \multicolumn{1}{l}{$\oast$}                  & \multicolumn{1}{l}{$\oast$}                  & \multicolumn{1}{l}{\stilltodo{$\star$}}                                                                      & \multicolumn{1}{l}{$\oast$}                                                                          & \multicolumn{1}{l}{$\oast$}                  & \multicolumn{1}{l}{$\oast$}                                                                       &         \multicolumn{1}{l}{\stilltodo{$\star$}}                                                                           \\ 
Amazon                                              & \multicolumn{1}{l}{$\oast$}                                                                        & \multicolumn{1}{l}{$\oast$}                                                                         & \multicolumn{1}{l}{$\oast$}               & \multicolumn{1}{l}{\stilltodo{$\star$}}                  & \multicolumn{1}{l}{$\oast$}                  & \multicolumn{1}{l}{$\oast$}                                                                      & \multicolumn{1}{l}{\stilltodo{$\bigcirc$}}                                                                          & \multicolumn{1}{l}{\stilltodo{$\bigcirc$}}                  & \multicolumn{1}{l}{$\oast$}                                                                       &      \multicolumn{1}{l}{}             \\ \bottomrule

\end{tabular}
\end{table}
We observe that Signal 
 only reports the collection of phone number (personal information) in their data safety section. 
In contrast, our experiment reveals that the app's user interfaces also collect location data (country name) and access data such as username, password, PINs. 
Furthermore, the app's source code uses system APIs that collect device IDs such as IP address and MAC address, personal information borrowed from Google such as client ID, email address, user ID, name, profile photo. 
It also collects audio, location, and browsing data. 
Our findings indicate that Signal collects device data such as SIM card information, but the data safety form does not have a separate category to report such data. 

We observe that Signal declares 70 permissions in its manifest file. 
A careful examination of these permissions reveals that their data safety section should include the collection of audio, contacts, location, messages, and calendar data. 
This confirms our hypothesis that \textit{Signal under-reports data collection in their data safety form}.
    
In contrast to Signal, Instagram\footnote{\href{https://www.instagram.com/}{https://www.instagram.com/}} reports the collection of \textbf{all} data categories in their data safety form. 
However, our results reveal that Instagram's user interfaces only collect financial information, specifically card number and CVV. 
The source code uses system APIs which collect device IDs such as IP address and MAC address, location, and personal information. 
Instagram also collects device data such as mobile network information, but the data safety form does not have a separate category to report such data.  

We extract the 40 permissions that Instagram declares in its manifest file. 
A careful examination of these permissions reveals that their data safety section should report collecting personal information, audio, contacts, location, and photos and videos. 
Consequently, the inclusion of other data categories in their existing form such as calendar, health data appears redundant, thus confirming our hypothesis that \textit{Instagram over-reports data collection in their data safety section.} \\

\begin{mdframed}[backgroundcolor=black!10,roundcorner=8pt]
\textbf{\textit{Finding 2.}} Our experiment reveals evidence of over- and under-reporting by Android app developers. We also identify the most inconsistently reported data categories (location, health and fitness, and calendar).
\end{mdframed}

Our prototype takes an average of 10.33 seconds per app to analyze the user interfaces. 
We manually analyzed all 362 records extracted from the user interfaces of the 20 apps, identifying 308 true positives, 15 false positives, 31 true negatives, and 10 false negatives. 
The resulting precision is \textbf{0.953}, and the recall is \textbf{0.968}, yielding an F1 score of \textbf{0.959}. We use an established static analysis tool, SootFX~\cite{sootfx}, to detect system API sources, which takes an average of 71.42 seconds to analyze the app source code. 

\section{Discussion}
\label{discussions}

We examine the data safety sections of the most popular apps across different domains and observe that many Android apps either over-report or under-report data collection. 
This emphasizes the importance of supporting developers better in accurately reporting data collection.
Our findings indicate that app developers struggle to accurately report data collection, despite the existence of tool support~\cite{matcha,privadoai,googlechecks}. 

One major reason for this issue could be the suboptimal design of Google's data safety form.  
While our experiment considers data collected via user interface
screens, system API calls in the app code, or manifest permissions as collected data, Google's definition of \textit{collected data} is abstract and includes many exemptions~\cite{dsscollection}. 
For example, one does not need to disclose the collection of anonymized data, although the effectiveness of common anonymization techniques is disputed in the digital privacy community~\cite{anonymize}. 
A study by Mozilla~\cite{mozilla} questioned Google's decision and suggested eliminating the definition of anonymized data from the data safety form. 
In contrast, pseudonymized data needs to be disclosed, while end-to-end encrypted data does not need reporting. 
This raises the question of how app developers determine if data has been properly anonymized or pseudonymized. 
Additionally, data processed locally on the user’s device and not sent off the device does not need to be disclosed either, which seems questionable, as substantial information might still be inferred from such data.

Another reason for under-reporting may be that the form lacks a separate category for data collected by third-party libraries used by the app. 
This could lead developers to believe they do not need to disclose data collected by these libraries~\cite{datalabels}, potentially giving users a false sense of privacy. 
This third-party collection information might only be available in the privacy policy of the app (if at all), which is often too long and verbose for users to read. 
Privacy-conscious users might want to know which data is collected by third-party analytics libraries used by the app, and adding this information is crucial towards enhancing transparency. 

After examining the data collected by real-world Android apps, we believe that \textbf{all} personal data collected by an Android app should be declared in the \textit{data collected} section of the data safety form, irrespective of whether it is pseudonymized or shared with third parties. 
The data shared with third parties can then be reported separately in the \textit{data shared} section of the form. 
Furthermore, the form should be reordered to follow a risk-based reporting of collected data. 
For example, directly identifiable personal data like \textit{device or other IDs} should be placed on top, followed by the less risky categories. 
A comparison of our results with the data safety section reveals that the data categories in the data safety form are not expressive enough to report all the data the system APIs collect from the device. 
Data categories in the form should be updated after conducting a thorough research of the system APIs that Android apps use to collect data. 
We make the identifier datasets open source\footnote{Available at \href{https://zenodo.org/records/13318828}{https://zenodo.org/records/13318828}} and public so that they can be used to enhance transparency and clarity in the data reporting documents. 
We further suggest the inclusion of the following new data categories:
\begin{enumerate}
\item \texttt{Device data}: Android APIs and other libraries collect some data from the device. 
This data can be partially identifiable and hence is different from the identifiable category of \textit{Device or other IDs}. 
We suggest including a new data type for such data. 
Examples of such data are SIM card information, bluetooth information and mobile network information.
\item \texttt{Session data}: Media players, advertising platforms and analytics libraries collect session data to analyse user behavior, which does not have a distinct data type in the data safety form.
\item Sub-types in the data category \texttt{Device or other IDs}: We observe \textit{Device or other IDs} to be a very coarse representation, and recommend distinguishing these IDs by their purposes. 
For example, IP address may be used for gaining access to location and installation ID may be used for advertising and analytics.
\end{enumerate}

Khandelwal et al.'s study~\cite{datalabels} noted a lack of educational resources and demos to aid app developers in completing the data safety section. 
Our findings indicate that app developers may be unaware of Google's guidelines for Android developers~\cite{perm}. 
These guidelines include a list of manifest permissions and system APIs that should directly correspond to specific data safety labels. 
For example, Signal (discussed in Section~\ref{experiment}) declares \textbf{70} permissions, many of which should directly translate to data safety labels. 
Nevertheless, Signal has the most under-reported data collection among the 20 apps discussed in Table~\ref{tab:dsscasestudy}. 

Furthermore, we observe that Google's guidelines for Android developers~\cite{perm} and Google's data collection exemptions for the data safety section~\cite{dsscollection} can be complementary, potentially causing confusion for app developers. 
According to the Android developers' guide, Signal's usage of permissions should translate to data safety labels. 
However, if the data collected via those permissions is encrypted in the source code, app developers might refer to Google's data safety exemptions and choose to not report it as part of the data safety section. 
Our observations highlight the need for standardized guidelines and rules before introducing more tools to automate the reporting process. \\

\begin{mdframed}[backgroundcolor=black!10,roundcorner=8pt]
\textbf{\textit{Takeaway.}} App developers struggle to accurately report data collection, either due to the suboptimal design of Google's data safety form, lack of standardized guidelines, or insufficient existing tool support.
\end{mdframed} 
\section{Threats to Validity}
\label{threats}
We next discuss the threats to the validity of our experiment, and how we sought to mitigate them.

\textbf{Datasets.} The accuracy of the results depends on how precisely one constructs the datasets. 
The degree of identifiability relies on human subjectivity and it may vary with different domains and compliance regimes. 
The proposed definition of data focuses on GDPR but modifying the datasets is sufficient to adapt the analysis to alternative compliance regimes.

Since certain keywords in the \textit{identifier keywords dataset} may have different meanings in different contexts, they may produce false positives. 
In early experiments, we observed that the UI analysis automatically labeled the keywords ``body'' and ``height'' as \textit{partially identifiable health and fitness data}. 
In reality, input fields in messaging and e-commerce apps use the keyword ``body'' when describing the body of a chat or an email, and many apps use the keyword ``height'' to describe the dimensions of an image, making these input fields \textit{context-dependent data} instead. 
The mislabeling initially occurred because we prioritized labeling risk 2 data (partially identifiable data) before moving on to risk 4 data (context-dependent data), causing these keywords to be labeled incorrectly. To address this issue, we adjusted the order of precedence levels of \mbox{risk 2} health data in different domains, reducing false positives by 0.5\%.  
We consider the remaining false positives to be insignificant. 
However, in the future, we aim to explore more advanced Natural Language Processing techniques similar to UiRef's disambiguation strategy~\cite{uiref} to further mitigate the risk of false positives. 

Despite the process of third-party library selection we outlined in Section~\ref{approach} and Table~\ref{tab:tpls}, the \textit{identifier API dataset} may miss out on some libraries. 
We will explore whether using library detection tools~\cite{libd, libradar, libdetection} can help us cover a wider range of libraries. 

\textbf{Prototype.} The UI analysis is currently limited to textual data collected by an Android app, but it can be broadened to include a wider range of UI widgets. 
It focuses on scanning the layout files of Android apps, and does not associate privacy-related input fields with their corresponding variables in the source code. 

For system API analysis, SootFX matches the method signatures in the call graph with the labeled methods present in the \textit{identifier API dataset}. 
Since it uses keyword matching to detect the presence of data sources including those from some third-party libraries, its analysis will not be able to handle code obfuscation in cases where the obfuscation renames also calls to these third-party APIs.

\textbf{Experiment.} We start our experiment with a manual analysis of the data safety sections of 70 Android apps, 10 per domain. 
We then focus on the 20 apps that seem to be over-reporting and under-reporting data collection.  
While the number of apps might appear limited, we ensure a comprehensive evaluation by including the most popular apps from different domains actively used by people. 
However, selecting top apps per domain could threaten external validity, so our findings may not be generalizable to other domains. 
\section{Conclusion}
\label{conclusion}

In this paper, we have introduced a multi-layered definition of privacy-related data to correctly report data collection in Android apps. This data definition focuses on GDPR's definition of personal data. We have further described how we created a dataset of privacy-sensitive data classes that researchers may use to classify inputs to Android apps. This dataset takes into account data collected both through the user interface and system APIs. 

We have conducted a case study to observe how data collection is reported via the data safety sections of 70 apps, and observed over- and under-reporting of collected data (\textbf{RQ1}). 
Additionally, we have developed a prototype to statically extract and label privacy-related data collected by an Android app, and compared its results with the data safety sections of 20 apps. 
Using the results from two \textit{Messaging and Social Media} apps (Signal and Instagram), we have discussed how app developers under-report and over-report data collection, respectively, and identified inaccurately reported data categories (\textbf{RQ2}). 
Our findings have shown that app developers struggle to accurately report data collection, either due to the suboptimal design of Google’s
data safety form, lack of standardized guidelines, or insufficient existing tool support.

We next aim to interview Android developers to understand their perceptions of collected data and evaluate whether existing tools assist them effectively in completing the data safety section. 
\begin{acks}
We thank Michael Schlichtig for his help and encouragement in conducting this study. We also thank the anonymous reviewers for their feedback and suggestions.
\end{acks}

\bibliographystyle{ACM-Reference-Format}
\bibliography{sample-base}

\end{document}